\documentclass[conference]{IEEEtran}

\usepackage[utf8]{inputenc}
\usepackage{amsmath}
\usepackage{cite}
\usepackage{multicol}
\usepackage{graphicx}
\usepackage{array}
\usepackage{multirow}
\usepackage{graphicx}
\usepackage{enumitem}
\usepackage{color}
\usepackage{nomencl}
\usepackage{etoolbox}
\usepackage{amssymb}
\usepackage{comment}
\usepackage{algorithm}
 \usepackage{algpseudocode}
\usepackage{mathtools}
\usepackage{bbold}
\usepackage{yfonts}
\usepackage{soul}
\usepackage{hyperref}
\usepackage{fnpct}
\usepackage{booktabs}
\ifCLASSOPTIONcompsoc \usepackage[caption=false,font=normalsize,labelfon
t=sf,textfont=sf]{subfig}
\else
\usepackage[caption=false,font=footnotesize]{subfi g}
\fi
\usepackage{tikz}
\usetikzlibrary{decorations.pathreplacing}
\usetikzlibrary{arrows,shapes,positioning}
\usetikzlibrary{patterns}
\usepackage{pgfplots}

\title{Iterative McCormick Relaxation for Joint Impedance Control and Network Topology Optimization}
\author{
\IEEEauthorblockN{Junseon Park$^{*,\S}$, Hyeongon Park$^*$, Rahul K. Gupta$^\S$\\
$^*$Pukyong National University, Busan, South Korea, $^\S$Washington State University, Pullman, WA, USA.}}
\makeatletter
\patchcmd{\@maketitle}
  {\addvspace{0.5\baselineskip}\egroup}
  {\addvspace{-1\baselineskip}\egroup}
  {}
  {}
\makeatother
\begin{document}
\maketitle
%%%
\begin{abstract}
Power system operators are increasingly deploying Variable Impedance Devices (VIDs), e.g., Smart Wires, and Network Topology Optimization (NTO) schemes for mitigating operational challenges such as line and transformer congestion, and voltage violations. This work aims to optimize and coordinate the operation of distributed VIDs considering fixed and optimized topologies. 
This problem is inherently non-linear due to power flow equations as well as bilinear terms introduced due to variable line impedance of VIDs. Furthermore, the topology optimization scheme makes it a mixed integer nonlinear problem. To tackle this, we introduce using  McCormick relaxation scheme, which converts the bilinear constraints into a linear set of constraints along with the DC power flow equations. We propose an iterative correction of the McCormick relaxation to enhance its accuracy.
% and NTO schemes while minimizing the total operating costs. Due to the control on impedance as well as the topology optimization, this optimization scheme result in mixed integer bilinear program. In this work, we explore the use of 
% The coordination and optimization of these technologies require solving non-linear and non-convex optimization schemes
% However, as the deployment of multiple GETs grows, effective coordination among them becomes essential to fully realize their potential benefits. This paper presents a co-optimization framework that models and coordinates NTO, VID, and DLR within a unified optimization scheme to alleviate network congestion and minimize operational costs. The NTO formulation is developed using a node-breaker model, offering finer switching granularity and improved operational flexibility. The inclusion of VIDs introduces nonlinear and non-convex relationships in the optimization problem. DLR takes into account of weather conditions, primarily wind speed and ambient temperature, enabling adaptive utilization of transmission capacity. 
The proposed framework is validated on standard IEEE benchmark test systems, and we present a performance comparison of the iterative McCormick method against the non-linear, SOS2 piecewise linear approximation, and original McCormick relaxation.
% , demonstrating its effectiveness under varying numbers and placements of impedance controllers. We perform sensitivity analysis with nonlinear and SOS2 formulation
% different settings of the McCormick relaxation constraints.
\end{abstract}
%%%%%%%%%%%%%%%%%%%%%%%%%%

% \section{Introduction}
% I will explain about McCormick Relaxation.
% %%%%%%%%%%%%%%%%%%%%%%%%%%
% \section{Problem Formulation}
% I will explain about Problem Formulation.

% \clearpage

\section{Introduction}
% \subsection{Motivation}
The power system is facing operational challenges such as power flow congestion and voltage quality issues \cite{liang2016emerging, NERC_report} due to massive interconnection of intermittent renewable energy resources (RERs), inverter-based resources (IBRs) \cite{lin2022pathways} as well as data center loads \cite{DOE_report_RA}. On the one hand, utilities are looking for different long-term solutions such as network expansion, transformer upgrade \cite{de2008transmission}, etc. On the other hand, 
% These developments introduce and introduces new challenges to the power system utilities to ensure the reliability of the power system. Conventional solutions to these challenges are to expand transmission capacity through new line construction or infrastructure upgrades \cite{lee2006transmission, de2008transmission}, deploy energy storage systems (ESS) \cite{denholm2010role, piansky2025optimizing}, etc. However, these approaches are often capital-intensive and time-consuming, requiring several years for planning, permitting, and commissioning.
several Grid Enhancing Technologies (GETs) have emerged as promising alternatives to improve network flexibility and mitigate congestion issues \cite{DOE_report}. These include Network Topology Optimization (NTO) \cite{xiao2018power}, Variable Impedance Devices (VIDs), also referred to as Smart Wire Devices (SWDs) \cite{sahraei2016computationally}, etc. 
% NTO seeks to optimize both the network topology and the breaker configurations, often represented by the node-breaker model, to enhance operational flexibility. Prior studies \cite{xiao2018power} have shown that NTO can significantly reduce operational costs compared to fixed-topology systems.  Meanwhile, VIDs (or SWDs) can modulate the line impedance, effectively rerouting power flows from overloaded lines to underutilized ones \cite{sahraei2016computationally}.
% Recognizing the potential of these technologies, the U.S. Department of Energy (DOE) has recently emphasized the importance of GET deployment to enhance grid reliability and flexibility \cite{DOE_report}. 

% NTO simultaneously optimizes network topology and breaker configurations, typically modeled using a node-breaker representation, to improve operational flexibility.
Prior work has demonstrated that NTO can significantly reduce operational costs relative to fixed-topology systems \cite{xiao2018power}, and improve operational flexibility. In contrast, VIDs, or SWDs, dynamically adjust line impedance to redistribute power flows from congested to lightly loaded transmission lines \cite{sahraei2016computationally}. VIDs are a type of flexible AC transmission system (FACTS), such as Thyristor Controlled Series Compensator (TCSC), Static Synchronous Series Compensator (SSSC), Unified Power Flow Controller (UPFC), etc., and can be modeled by variable line impedance. The variable impedance/reactance allows flexibility in the transmission network to re-route the power flows and help avoid congestion and reduce costs.
% Recognizing the potential of these technologies, the U.S. Department of Energy (DOE) has emphasized the deployment of GETs as a critical strategy for improving grid reliability and flexibility \cite{DOE_report}.
% However, most existing methods address these technologies independently, without considering their interactions or coordination in congestion management. To address this gap, this work proposes a co-optimization framework that jointly models and coordinates NTO, DLR, and distributed VIDs, enabling synergistic operation to alleviate transmission congestion and improve overall system efficiency.

% \subsection{Existing literature} 
% [TODO: With respect to modeling and combinations of these technologies]
In the existing literature, the optimization of distributed VIDs across has been widely investigated. Incorporating VIDs into operational planning introduces additional nonlinearities in both the optimal power flow (OPF) and switching problems, as VIDs are typically modeled through variable line impedance \cite{sahraei2016computationally, nikoobakht2018smart}. Different methods have been proposed to tackle the nonlinearities. For example, the work in \cite{sahraei2016computationally} proposes a mixed integer reformulation; however, the controls are limited to the maximum and minimum operation setpoints of the VIDs. A similar strategy was used in \cite{nikoobakht2018smart} for the placement of the SWDs to facilitate integration of large-scale wind power plants. Overall, these methods are focused on the planning of the SWDs. The work in \cite{pohl2024coordination} developed a scheme for the operation of SWDs using a sensitivity factor approach, but they are not applicable to different topologies.

In this work, we focus on the joint optimal operation of the VIDs with base and optimized topology. 
First, the network topology and bus-bar splitting are modeled using the node-breaker representation, which enables detailed switching decisions while maintaining network connectivity and operational feasibility. Then, VIDs are incorporated to provide additional flexibility by allowing controlled adjustments of line reactances within predefined operating limits, thereby redistributing power flows and alleviating congestion. The DC power flow model is adopted to represent power flow physics, extended to account for the variable line reactances introduced by the VIDs. 

As this problem has bilinear constraints, we propose a McCormick relaxation-based approach to linearize the bilinear constraints in the DC-OPF and NTO problems. However, this scheme usually results in large errors due to the McCormick relaxation, especially when the allowable change in the network impedance is large, for example, 30-50\% of the rated value. 
For enhancing the accuracy of the McCormick relaxation approach, we propose an iterative McCormick method, where the bounds on the maximum allowable range in the impedance is changed in steps. 
% Then, we reformulate the problem using two versions of McCormick relaxation.
The key contribution of this work lies in linearization techniques using the iterative McCormick relaxation approach and its performance comparison with respect to the nonlinear and mixed integer schemes.

% the development of a co-optimization scheme that jointly considers NTO, DLR, and VID technologies within a single, integrated decision-making framework. This unified approach enables a systematic assessment of the interactions and combined benefits of multiple GETs under diverse operating and weather conditions. Through this framework, we can evaluate how coordinated control of these technologies enhances network flexibility, increases renewable energy accommodation, and improves overall system reliability, outcomes that are difficult to achieve when each technology is optimized independently.

The paper is organized as follows. Section~\ref{sec:methodology} presents the problem formulation. Section~\ref{sec:co-optimization} presents the relaxation using the McCormick method and SOS2 approximation. Section~\ref{sec:numerical_sim} presents the simulation setup and results on different IEEE testcases, and finally, Section~\ref{sec:conclusion} concludes the main contribution of the presented work.
%%%
\section{Problem Formulation}
\label{sec:methodology}
We consider a power transmission network consisting of $N_b$ buses, $N_g$ generators and connected with several VIDs. The objective is to optimize the operation of generators, load-shedding, distributed VIDs and the network topology. The optimization objectives and the constraints are defined below.
\subsection{Objective function}
The proposed optimization aims to minimize the total system operating cost, which consists of (i) generation cost $(C^{Gen})$ and (ii) load-shedding cost $(C^{LS})$, defined as follows.
\begin{subequations}
% \begin{align}
%    C^{Obj} =  C^{Gen} + C^{LS} %+ C^{VID} 
%    \label{total}
% \end{align}
% % Objective function \eqref{total} aims to minimize the total operating cost of the power system. The objective function consists of three cost function elements: the generation cost of the generator, the load shedding cost, and Variable impdedance Operation Cost.  
% Each of these individual costs are defined as follows.
\subsubsection{Generation cost}
They are modeled as a quadratic function, as shown in Equation \eqref{gen_cost}, where the net generation is the sum of the generation produced at the two busbars.

\begin{small}
\begin{align}
& C^{Gen} = \sum_{g = 1}^{N_g}({c_{g,2}({P_{g, 1} + P_{g, 2}})^2 + c_{g,1}({P_{g, 1} + P_{g, 2}}) + c_{g,0}}) \label{gen_cost}
\end{align}
\end{small}
where, $c_{g, 0}$, $c_{g,1}$ and $c_{g, 2}$ are the cost parameters of the generators and $P_{g,1}$ and $P_{g,2}$ refer to the generator units at busbar 1 and 2, respectively.
%%%%%%%%%%%%%
\subsubsection{Load shedding cost}
When transmission capacity constraints fail to meet the total load demand, load shedding may occur. Multiplying the load shedding power by a high-cost $VOLL$ minimizes the load shedding cost.  The load shedding cost can be calculated as shown in Equation \eqref{load curtail}.
\begin{align}
& C^{LS} =  VOLL \sum_{b = 1}^{N_b}(\sum_{d \in D_b}({P_{b,d}^{max} - (P_{{b,d}, 1} + P_{{b,d}, 2})})) \label{load curtail}
\end{align}
where, VOLL refers to the value of loss of load, $P_{b,d,1}$ and  $P_{b,d,2}$ are the demand at busbars 1 and 2 of bus $b$. The symbol $P_{b,d}^{max}$ refer to the maximum demand at bus $b$. $D_b$ represents a set containing the indices of the demand per bus $b$.
%%%%%%%%%
% \subsubsection{Variable Impedance Operation Cost}
% The cost for using VID i is the product of the cost required to change the unit susceptance and the amount of susceptance change. 
% \begin{align}
% & C^{VID} = \sum_{l = 1}^{NL}{\lambda_{l}|\Delta b_l|} \label{XCDF_cost}
% \end{align}   
\end{subequations}

% grid model and the models of different GETs, which will be used later for the co-optimization framework in Sec.~\ref{sec:co-optimization}.
%% This study employed the following three methodologies to reduce line congestion: Dynamic Line Rating (DLR), Network Topology Optimization (NTO), and Variable Impedance Devices (VID).\\
% DLR is a methodology that updates line transmission capacity by incorporating real-time weather data. In contrast, Static Line Rating (SLR) does not incorporate real-time weather data and instead fixes transmission capacity based on time variations. This method is typically set conservatively, thus not allowing transmission of more power in most cases even when it is available. Furthermore, as extreme weather events increase due to global warming, this approach enables reliable operation of the power system in advance.\\
% NTO introduces a Node Breaker model to alter the power system topology. Unused lines can be disconnected, and lines, generators, and loads can be optimally selected and operated with busbars, enabling more flexible power system operation compared to traditional fixed topology models. \\
% Variable Impedance Devices model is a methodology that reduces line congestion and enables efficient power system operation by adjusting the impedance within a certain range for congested lines. Changing impedance changes the power flow. This allows power to be sent more to lines without congestion by reducing their impedance, while increasing impedance on congested lines to reduce load shedding. This reduces power system operating costs. \\

\subsection{Constraints}
The constraint set consists of a power flow model, constraints on the network topology, constraints on the generators and load-shedding. These are described below.
\subsubsection{Transmission Grid Model}
We model the transmission network using the DC power flow approximation \cite{stott2009dc}, which assumes that (i) the angle of the voltage difference is small, (ii) the network is reactive, and (iii)  the voltage magnitudes are close to 1~per unit.
% and the transformer's tap settings are neglected. When using DC Power Flow for optimization, Since there is no need to consider reactive power or nonlinear terms, solutions can be obtained easily and quickly. 
Let the symbol $b_l$ represent the susceptance of line $l$, $\theta_{l,fr}$ and  $\theta_{l,to}$ denote the voltage angle for line $l$ at ''from'' and ''to'' ends.
Then, the flow in line $l$, $P_l$ can be expressed using DC power flow model as
\begin{align}
& P_{l} = b_{l}(\theta_{l, fr} - \theta_{l, to}). \label{DC}
\end{align}
% Here, $b_{l} = \bar{b}_{l}$, i.e., when impedances are not controlled.
%%%%%%%%%%%%%%%%%%%%%%%%%%%
%%%%%%%%%%%%%%%%%%%%%%%%%%%
\subsubsection{Network Topology Optimization Model}
\begin{figure}[!htbp]
    \vspace{-0.5em}
    \centering
    \includegraphics[width=0.9\linewidth]{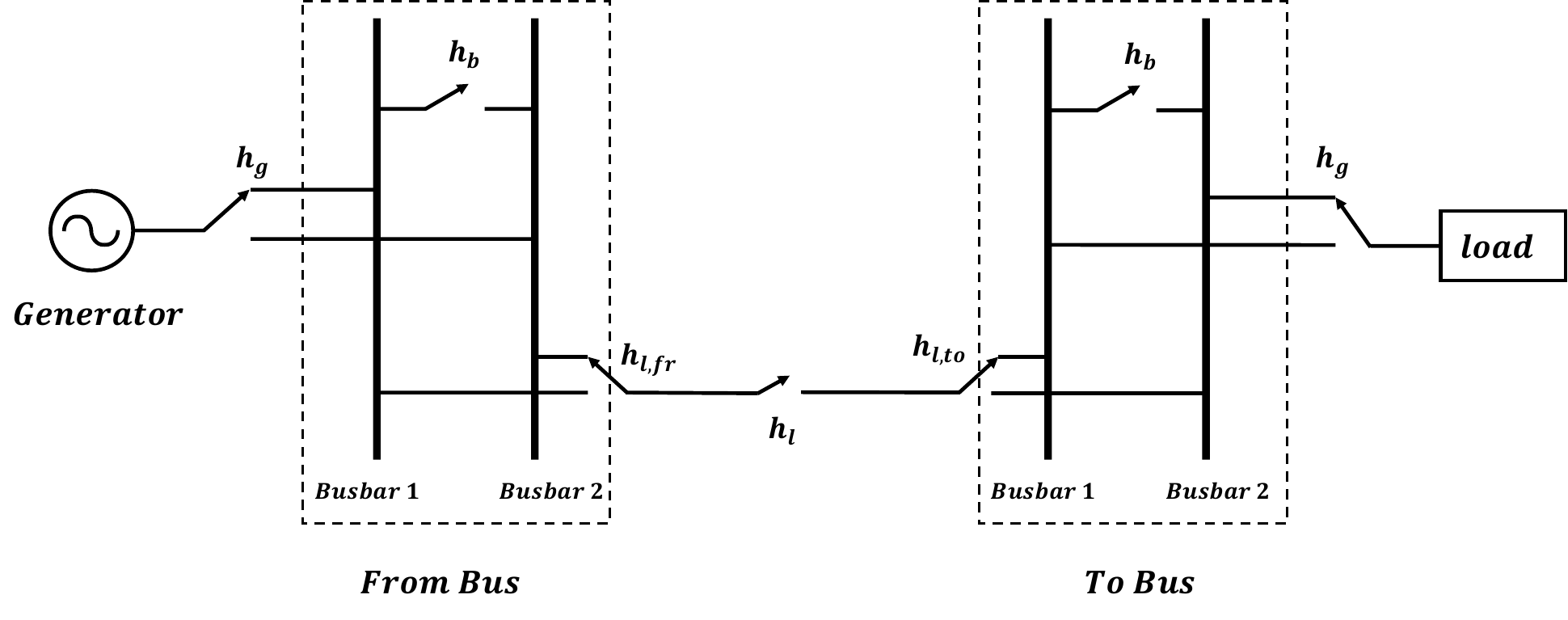}
     \vspace{-1em}
    \caption{Generalized breaker-and-half model.}
    \label{fig:nodebreaker}
    \vspace{-1em}
\end{figure}

% NTO involves optimizing the configuration of the power network through optimal transmission switching and by leveraging existing switching elements within substations, such as circuit breakers (CBs). The substation CBs enable reconfiguration of substation connections to the rest of the power system, thereby influencing the operational flexibility, reliability, and security of the grid. 
% Substations can employ various bus configurations that require different numbers of circuit breakers and exhibit distinct reliability characteristics. Among these, the breaker-and-a-half configuration is commonly used in high-voltage substations due to its favorable balance between reliability and operational flexibility \cite{atanackovic1999reliability}.
% In this work, we assume that the node-breaker (NB) model is adopted at each bus in the power network. Figure~\ref{fig:nodebreaker} shows the generalized breaker-and-half model with line switching. As shown in the figure, this model adds additional flexibility that the power system could benefit from, for example, using the NB model, the busbars of the ``from" and ``to" buses can be either closed or open, and also the connection to the loads and generators can be connected to either of the busbars. 
NTO identifies optimal network topology by utilizing controllable switching devices, including transmission switches and substation circuit breakers (CBs). Substation CBs enable reconfiguration of internal bus arrangements and their interfaces with the rest of the network, thereby affecting grid flexibility, reliability, and security. Various bus configurations are used in practice, each requiring different numbers of breakers and offering distinct reliability characteristics; among them, the breaker-and-a-half scheme is common in high-voltage substations due to its favorable balance of reliability and operational flexibility \cite{atanackovic1999reliability}.
In this work, we modeled the substations using the node-breaker (NB) representation. Figure~\ref{fig:nodebreaker} shows a generalized breaker-and-a-half layout with line-switching capability. This NB model provides additional reconfiguration options, allowing the “from’’ and “to’’ busbars to be opened or closed independently and enabling loads and generators to connect to either busbar.
% extends the OPF problem by not only considering power flow with a fixed network topology, but also by optimizing possible changes when the topology can vary.The generalized NTO Model is shown in Fig.~\ref{fig_model}. 

% The way to change the network topology is by using switching devices such as circuit breakers. There are four components that change the topology: busbars, lines, generators, and loads. Each substation consists of two busbars, which can be merged into a single bus or split into two separate buses. For lines, disconnecting unnecessary lines enables more flexible operation of the power system. Finally, generators and buses can be connected to either Bus 1 or Bus 2. 
Considering the switching capability of the busbars and the lines, the NTO model can be mathematically modeled using busbar binary variables $h_b$ \cite{7038226, xiao2018power}: 
\begin{subequations}    
\label{eq:NTO_equations}
\begin{align}
& -\theta^{max} (1 - h_b) \leq \theta_{b, 1} - \theta_{b, 2} \leq \theta^{max} (1 - h_b)\quad \forall b, \label{max_voltage}
\end{align}
where $\theta_{b,1}$ and $\theta_{b,2}$ are the voltage angles at busbars 1 and 2 and $\theta^{max}$ is the maximum difference of angle between the busbars.
In \eqref{max_voltage}, when busbars are merged, the difference in voltage angle phase between busbars must be identical, and when split, it must not exceed the maximum angle phase. 

The generators connected at each busbar can be modeled using binary variables $h_g$, and can be expressed as
\begin{align}
& (1-h_g)P_g^{min} \leq P_{g, 1} \leq (1 - h_g)P_g^{max} \quad \forall g \label{Pgmax_1}\\
& h_gP_g^{min} \leq P_{g, 2} \leq h_gP_g^{max} \quad \forall g. \label{Pgmax_2}
\end{align}
% where, $P_{g,1}$ and $P_{g,2}$ are the generator setpoint at busbars 1 and 2 at generator $g$. 
The symbols $P_g^{min}$ and $P_g^{max}$ denote the minimum and maximum limits.
Eqs. \eqref{Pgmax_1} and \eqref{Pgmax_2} model when a generator is connected to busbar 1 or at busbar 2, and vice versa.   

Similarly, the connection and controllability of the demand at each of the busbars are modeled using binary variable $h_d$ using \eqref{Pdmax_1} and bounded by \eqref{Pdmax_2}
\begin{align}
& 0 \leq P_{b,d, 1} \leq (1 - h_d)P_{b,d}^{max} \quad \forall b \label{Pdmax_1}\\
& 0 \leq P_{b,d, 2} \leq h_dP_{b, d}^{max} \quad \forall b. \label{Pdmax_2}
\end{align}
% We also consider that the load can be curtailed \eqref{Pdmax_2} 
% % \eqref{Pdmax_2} model the curtailability of the demand as
% \begin{align}
% % & 0 \leq P_{d, 1} \leq (1 - h_d)P_d^{max} \quad \forall d \label{Pdmax_1}\\
% & 0 \leq P_{d, 2} \leq h_dP_d^{max} \quad \forall d \label{Pdmax_2}
% \end{align}
% , and can be expressed as

The limit on the power-flow for each line and the switching of the lines are expressed as
\begin{align}
 -(1 - h_{l, e})P_l^{max} &\leq P_{l, e, 1} \leq (1 - h_{l, e})P_l^{max} && \forall l, e \label{Plmax_1}\\
 - h_{l, e}P_l^{max} &\leq P_{l, e, 2} \leq h_{l, e}P_l^{max} && \forall l, e \label{Plmax_2}\\
 -h_{l}P_l^{max} &\leq P_{l, e, 1} \leq h_{l}P_l^{max} &&\forall l, e \label{hl_max}\\
 h_{l, e} &\leq h_{l} &&\forall l, e \label{hl}\\
 P_l &= P_{l, e, 1} + P_{l, e, 2} && \forall l, e \label{sumPl}
\end{align} 
where $h_{l,e}$ refer to binary variable representing whether line $l$ is on or off. $e = \{``\text{from}", ``\text{to}"\}$ refer to the ends of the line. $P_l^{max}$ defines the maximum capacity of line $l$. $P_{l, e, 1}$ and $P_{l, e, 2}$ refer to power flow of line $l$ at the end side of $e$ through busbar 1 and busbar 2 respectively. 
% $P_l$ indicates the power flow of line $l$.
% where, Eqs. \eqref{Plmax_1} and \eqref{Plmax_2} represent transmission line constraints: If a line is connected to busbar 1, the power flowing to the others must be zero, and vice versa. Conditions \eqref{hl_max} and \eqref{hl} represent the state of the line, indicating that power cannot flow when the line is open. Constraints \eqref{sumPl} represents the sum of power flowing from the busbar, requiring that power flowing from ‘from’ to ‘to’ must be conserved. 
Eqs. \eqref{Plmax_1} and \eqref{Plmax_2} define the transmission line constraints: when a line is connected to busbar 1, the power flow to the other busbars must be zero, and vice versa. Eqs. \eqref{hl_max} and \eqref{hl} specify the line status, ensuring that no power flows when the line is open. Constraint \eqref{sumPl} enforces power conservation by requiring that the total power flowing out of a busbar equals the power flowing from the ‘from’ bus to the ‘to’ bus.

The line switching constraints are expressed using the big-$M$ method, given by \eqref{Pf_NTO}, \eqref{theta_max_1}, and \eqref{theta_max_2} 
\begin{align}
   & -M(1 - h_l) \leq  b_{l}(\theta_{l, fr} - \theta_{l, to}) - P_l \leq M(1 - h_l) \; \forall l \label{Pf_NTO}\\
 & -h_{l, e}\theta^{max} \leq  \theta_{l, e} - \theta_{l, e, 1} \leq h_{l, e}\theta^{max} \quad \forall l, e \label{theta_max_1}\\
 &  -(1 - h_{l, e}\theta^{max}) \leq \theta_{l, e} - \theta_{l, e, 2} \leq h_{l, e}\theta^{max} \quad \forall l, e. \label{theta_max_2}
\end{align}
% Constraints \eqref{Pf_NTO}, \eqref{theta_max_1}, and \eqref{theta_max_2} describe the line flow constraints. 

The binary variables for busbars, generators, demand connections, and lines are connected through Eqs. \eqref{hg1}, \eqref{hd1}, and \eqref{hle1}.
% represent the constraints when the busbars at each substation are connected. 
When the busbars at each substation are connected, it is not necessary to consider the binary variable determining the connection of the generator, load, and end of the line.
\begin{align}
& h_b + h_g \leq 1 \quad \forall b \; g \in G_b \label{hg1}\\
& h_b + h_d \leq 1 \quad \forall b \; d \in D_b \label{hd1}\\
& h_b + h_{l, e} \leq 1 \quad \forall b, e \; l \in LF_b \; or \; l \in LT_b. \label{hle1}
\end{align}
% [\textcolor{red}{@Junseon: please verify if some variables are undefined, and define them in the text if needed ----- Completed work: Can you please check?}]
Finally, the constraints \eqref{balance1} and \eqref{balance2} express the power balance constraints, i.e., the difference between generation and demand is equal to the sum of power flowing through the lines. The symbol $G_b$ refers to a set containing the indices of the generator per bus $b$. $LF_b$ and $LT_b$ denote set of lines flow "from" bus $b$ and "to" bus $b$, respectively.
\begin{align}
& \sum_{g \in G_b}P_{g, 1} - \sum_{d \in D_b}P_{b,d, 1} - \sum_{l \in LF_b}P_{l} + \sum_{l \in LT_b}P_{l} = 0 \quad \forall b \label{balance1}\\
& \sum_{g \in G_b}P_{g, 2} - \sum_{d \in D_b}P_{b,d, 2} - \sum_{l \in LF_b}P_{l} + \sum_{l \in LT_b}P_{l} = 0 \quad \forall b \label{balance2}
\end{align}
\end{subequations}
\subsubsection{VID Model}
% Variable Impedance Devices model is a type of flexible AC transmission system (FACTS). They are modeled 
% are several kinds of FACTS devices, including Thyristor Controlled Series Compensators (TCSCs), Static Synchronous Series Compensators (SSSCs), and Unified Power Flow Controllers (UPFCs), which are devices that can change the impedance in a line\cite{pohl2024coordination}. 
% However, FACTS devices are large in scale and are expensive to install and produce. 
VIDs can be mathematically modeled as variable susceptance devices \cite{sahraei2016computationally, nikoobakht2018smart, 8403390} as
\begin{subequations}
\label{eq:VID}
\begin{align}
    b_l & = \bar{b}_{l} + \Delta b_l \\
    -r \bar{b}_l & \leq \Delta b_l \leq r \bar{b}_l \quad \forall l \label{rangeb}
\end{align}
\end{subequations}
% \begin{subequations}
% \label{XCDF}
% \begin{align}
%  & P_{l} = (\bar{b}_{l} + \Delta b_l) (\theta_{l, fr} - \theta_{l, to}) \quad \forall l \label{OPF_XCDF} \\ 
%  &\begin{aligned}
%       -M(1 - h_l) \leq & (\bar{b}_{l} + \Delta b_l)(\theta_{l, fr} - \theta_{l, to}) - P_l \\ 
%       \leq & M(1 - h_l) \; \forall l. \label{NTO_XCDF}
%  \end{aligned}
% \end{align}
% Here, \eqref{OPF_XCDF} applies to the fixed topology case, whereas \label{NTO_XCDF} used when considering network topology in the optimization.
% and \eqref{NTO_XCDF} are forms of Equations \eqref{DC} and \eqref{Pf_NTO} with $\Delta b_l$ added. Equation \eqref{OPF_XCDF} can be used for fixed topology optimization, while Equation \eqref{NTO_XCDF} can be used when considering network topology in the optimization. 
where, $\bar{b}_l$ denotes the nominal susceptance and $\Delta b_l$ is a variable allowing deviations in the line susceptance is bounded by a factor $r \leq 1$ of nominal susceptance $\bar{b}_l$.
% % from $b_l$ to $b_l + \Delta b_l$, line congestion can be reduced. 
% \begin{align}
% & -r \bar{b}_l \leq \Delta b_l \leq r \bar{b}_l \quad \forall l \label{rangeb}
% \end{align}
% \end{subequations}
% The devices that change impedance cannot change it infinitely; 
% representing that each device has its own limitations, so a range must be specified. 
% Constraint \eqref{rangeb} represents the upper bound and lower bound within the range that susceptance can be changed. 
% The work in \cite{nikoobakht2018smart} addresses the nonlinearity due to the VID constraints by introducing auxiliary variables. In this work, we use Gurobi as solver which is capable to solve the bilinear problem using McCormick Envelopes.
%%%%%%%%%%%%%%%%%%%%%%%%%
\subsection{Final Optimization Problem}
The final optimization can be formulated as
\begin{equation}
\label{eq:optimization_problem_final}
\begin{aligned}
\min \quad &  C^{Obj} \;=\; C^{Gen} + C^{LS}, \\
\text{subject to} \quad 
&\eqref{eq:NTO_equations},\eqref{eq:VID}.
\end{aligned}
\end{equation}

Note that the optimization problem in \eqref{eq:optimization_problem_final} is non-linear because of the bilinear variables $\Delta b_l(\theta_{l,fr} - \theta_{l,to})$ due to VIDs when \eqref{eq:VID} is substituted in \eqref{Pf_NTO} and \eqref{DC} with $\Delta b_l $ and $\theta_{l,fr}, \theta_{l,to}$ being variables. Therefore, we propose a linearization approach using McCormick Relaxation as described below.
\section{McCormick Relaxation and SOS2 Approximation}
\label{sec:co-optimization}
% We formulate a co-optimization problem that jointly coordinates the operation of multiple GETs, aiming to minimize the total operating cost while satisfying the respective constraints of each technology. The objective function and its cost components are described below.
%%%%%

\subsection{McCormick Relaxation Model}
\subsubsection{McCormick Relaxation (Non-iterative)}
% The optimization problem in \eqref{eq:optimization_problem_final} is non-linear because of the bilinear constraint of VID when \eqref{eq:VID} is substituted in \eqref{DC} and \eqref{Pf_NTO}. 
We relax bilinear terms, $\Delta b_l(\theta_{l,fr} - \theta_{l,to})$,  by defining 
% $\Delta b_l \times \theta_{l,fr}$ and $\Delta b_l \times \theta_{l,to}$  
auxiliary variables, then applying McCormick envelopes \cite{mitsos2009mccormick}. These McCormick envelopes transform the bilinear constraints to a set of linear constraints by defining upper and lower bounds on each of the variables in the bilinear term. We define upper and lower bounds on the variables involved in the bilinear constraint, i.e., $[b_l^{min}, b_l^{max}]$ and $[\Delta \theta_l^{min}, \Delta \theta_l^{max}]$, then the McCormick constraint set is defined as follows. We define a new variable $w_{l} = \Delta b_{l}\Delta\theta_{l}$ where $\Delta\theta_{l}  =  \theta_{l, fr} - \theta_{l, to} $. Then, the McCormick relaxation is given as \eqref{mc_1}-\eqref{mc_4}:
\begin{subequations} 
\label{eq:mccormick}
\begin{align}
% & \Delta\theta_{l}  =  \theta_{l, fr} - \theta_{l, to} \label{equal_theta}\\
% & w_{l} = \Delta b_{l}\Delta\theta_{l} \label{ax_var} \\
& w_{l} \geq b_l^{min}\Delta\theta_{l} + \Delta b_{l}\Delta\theta_l^{min} - b_l^{min}\Delta\theta_l^{min}  \quad \forall l \label{mc_1}\\
& w_{l} \geq b_l^{max}\Delta\theta_{l} + \Delta b_{l}\Delta\theta_l^{max} - b_l^{max}\Delta\theta_l^{max}  \quad \forall l \label{mc_2}\\
& w_{l} \leq b_l^{min}\Delta\theta_{l} + \Delta b_{l}\Delta\theta_l^{max} - b_l^{min}\Delta\theta_l^{max}  \quad \forall l \label{mc_3}\\
& w_{l} \leq b_l^{max}\Delta\theta_{l} + \Delta b_{l}\Delta\theta_l^{min} - b_l^{max}\Delta\theta_l^{min}  \quad \forall l. \label{mc_4}
\end{align}
\end{subequations}

% \textcolor{blue}{\eqref{equal_theta} definedthe voltage phase angle difference between the ‘from’ and ‘to’ voltage of the line as a variable.\eqref{ax_var} defines the bilinear term $\Delta b_{l}\Delta\theta_{l}$ as a single auxiliary variable $w_{l}$. 
% Constraints \eqref{mc_1}-\eqref{mc_4} express the relaxation of the bilinear problem into four linear inequality constraints.

\subsubsection{Iterative McCormick Relaxation} 
The accuracy of the McCormick relaxation scheme is strongly influenced by the range of bounds associated with the variables in the McCormick equations. As it will be demonstrated in the Results section, the objectives obtained using the McCormick relaxation method exhibit a deviation from the costs computed by a nonlinear solver. To address this discrepancy, we propose an iterative McCormick relaxation technique, formally described in Algorithm~\ref{alg:alg1}. This algorithm applies McCormick relaxation in small steps, so that the accuracy is not compromised.
% The accuracy of the McCormick relaxations scheme depends on the range of the bounds of the variable involved in the McCormick equations. As it will be shown later the results section, the objectives obtained using McCormick relaxation method has some error compared to the costs obtained using non-linear solver. So, we propose an iterative McCormick relaxations technique - it is defined in Algorithm~\ref{alg:alg1}.
% where the value of $r$ is increase in steps. For example, if we want to simulate for $r = 0.2$, the McCormick relaxation is solved in steps of $r = 0.05$, then reset susceptance as $\hat{b}_l  = b_l + \Delta b_l$, then solve again till we reach $r = 0.2$. This process is repeated for any $r$.
\begin{algorithm}[H]
\caption{Iterative McCormick Relaxation with Incremental $r$ Updates}
\begin{algorithmic}[1]
\Require Target relaxation level $r_{\text{final}}$, step size $\Delta r$, initial susceptances $\{b_l\}$
% \State $r \gets 0$
\State Initialize $\hat{b}_l \gets b_l$ for all lines $l$
\While{$r < r_{\text{final}}$}
    \State Solve \eqref{eq:optimization_problem_final} using McCormick relaxation \eqref{eq:mccormick} for  $r = \Delta r$
    \State Compute susceptance update $\Delta b_l$
    \State Update susceptance: $\hat{b}_l \gets b_l + \Delta b_l$ for all $l$
    \State $r \gets r + \Delta r$
\EndWhile
\State \textbf{return} final solution at $r_{\text{final}}$
\end{algorithmic}
\label{alg:alg1}
\end{algorithm}

\subsection{SOS2 Model}
\label{sec:SOS2_model}
The bilinear term $b_l \Delta\theta_l$ in \eqref{DC} and \eqref{Pf_NTO} can also be approximated by a two-dimensional piecewise-linear model using special ordered sets of type~2 (SOS2). For each line $l$, the range $[b_l^{\min}, b_l^{\max}]$ and $ [\Delta\theta_l^{\min}, \Delta\theta_l^{\max}]$ are divided into $N_b^{\text{grid}}$ and $N_\theta^{\text{grid}}$ number of interpolation points, denoted by $\hat b_{l,i}$ and $\Delta\widehat{\theta}_{l,j}$.
% divide this into $N_b^{\text{grid}}$ and $N_\theta^{\text{grid}}$ parts. 
% \begin{align}
% $b_l^{\min} \le b_l \le b_l^{\max}$ and $\Delta\theta_l^{\min} \le \Delta\theta_l \le \Delta\theta_l^{\max}$, we
% construct breakpoints
% \begin{align}
%     &\hat b_{l,i} \in [b_l^{\min}, b_l^{\max}], 
%     && i = 1,\dots, N_b^{\text{grid}}, \\
%     &\Delta\widehat{\theta}_{l,j} \in [\Delta\theta_l^{\min}, \Delta\theta_l^{\max}], 
%     && j = 1,\dots, N_\theta^{\text{grid}},
% \end{align}
SOS2 constraints are constructed by defining continuous variables $\lambda_{l,i,j}, \alpha_{l,i}, \beta_{l,j}$ such that
\begin{subequations}
\label{eq:sos2_lambda_short}
\begin{align}
    &\sum_{i=1}^{N_b^{\text{grid}}}\sum_{j=1}^{N_\theta^{\text{grid}}}
      \lambda_{l,i,j} = 1;~ \alpha_{l,i} = \sum_{j} \lambda_{l,i,j};~
     \beta_{l,j} = \sum_{i} \lambda_{l,i,j}.
\end{align}
\end{subequations}
The vectors $(\alpha_{l,1},\dots,\alpha_{l,N_b^{\text{grid}}})$ and $(\beta_{l,1},\dots,\beta_{l,N_\theta^{\text{grid}}})$ are modeled as SOS2 sets, ensuring that only two adjacent breakpoints are active along each axis. In our implementation, $\lambda_{l,i,j}$, $\alpha_{l,i}$, and $\beta_{l,j}$ are modeled as continuous variables, while the adjacency structure of the piecewise–linear model is imposed through the SOS2 constraints; YALMIP\footnote{\url{https://yalmip.github.io/command/sos2/}} creates an SOS2 set and automatically adds the binary variables and linear constraints required to enforce the adjacency structure.

Using the interpolation weights, the line susceptance, angle difference, and an auxiliary variable $\tilde w_l$ that approximates the product $b_l \Delta\theta_l$
are defined as
\begin{subequations}
\label{eq:sos2_reconstruction}
\begin{align}
    & b_l = \sum_{i,j} \lambda_{l,i,j} \,\hat b_{l,i}, \quad
    \Delta\theta_l = \sum_{i,j} \lambda_{l,i,j} \,\widehat{\Delta\theta}_{l,j}, \\
    & \tilde w_l = \sum_{i,j} \lambda_{l,i,j} \,\hat b_{l,i}\,\widehat{\Delta\theta}_{l,j}.
\end{align}
\end{subequations}

Finally, the SOS2 approximation replaces the bilinear flow relation in \eqref{Pf_NTO} through
\begin{subequations}
\label{eq:sos2_flow_constraints}
\begin{align}
   &-M (1 - h_l) \le \tilde w_l - P_l \le M (1 - h_l), \label{eq:sos2_bigM}\\
   &-P_l^{\max} \le P_l \le P_l^{\max}, \label{eq:sos2_thermal}
\end{align}
\end{subequations}
where $M$ is a sufficiently large constant. When $h_l = 1$, \eqref{eq:sos2_bigM} enforces $P_l = \tilde w_l$, yielding a piecewise–linear approximation of $b_l \Delta\theta_l$. When $h_l = 0$, the constraint is relaxed, and the line flow is handled through the NTO constraints in \eqref{eq:NTO_equations}.

\section{Numerical Simulations}
\label{sec:numerical_sim}
\subsection{Simulation Setup}
We simulate our framework on four different IEEE benchmark test systems, which are \texttt{Case300}, \texttt{\detokenize{Case588_sdet}}, \texttt{Case1354pegase}, and \texttt{Case1888rte} \cite{babaeinejadsarookolaee2019power}, and will be described in the results section. We used the dataset from the MATPOWER testcases for the simulations.

For each testcase, we set the minimum output of all generators to zero in order to avoid situations where a generator’s minimum output would exceed the load at its corresponding bus.
To simulate the congestion scenario, where the control of VIDs and NTO would be beneficial, we artificially reduce the capacity of the transmission lines. This is done by first solve the OPF for the nominal case, then 
% optimizing the problem without transmission capacity constraints, the power flows on each line can be obtained. 
we set the capacity of the lines to be 80\% of the power-flows obtained in the nominal case. For the simulations with VIDs,we allow the susceptance change factor \eqref{rangeb}, $r$ is set from 0 to 0.5, allowing each line susceptance to vary up to 50\% of their nominal values.
% For the simulations with VIDs, we allow the susceptance change factor \eqref{rangeb}, $r$ to 0.1, allowing each line susceptance to vary by 10\%. 
The cost of load shedding (VOLL) is set to 2,000~$\$$/MWh.

\subsection{Results}
The results of the proposed iterative-McCormick relaxation is presented and compared against three different formulations: Nonlinear, McCormick, and SOS2. 
The performance are compared with respect to the achieved cost, computation time, and accuracy with respect to the nonlinear case. We use Gurobi 12.0\footnote{\url{https://www.gurobi.com}} as the optimization solver for all the cases.
% model for each IEEE test system, based on increasing the susceptance change factor for each model: Nonlinear, McCormick(Original), Iterative McCormick and SOS2.}
\subsubsection{Case300 System}
\begin{figure}[!htbp]
    \vspace{-1em}
    \centering
    \includegraphics[width=0.88\linewidth]{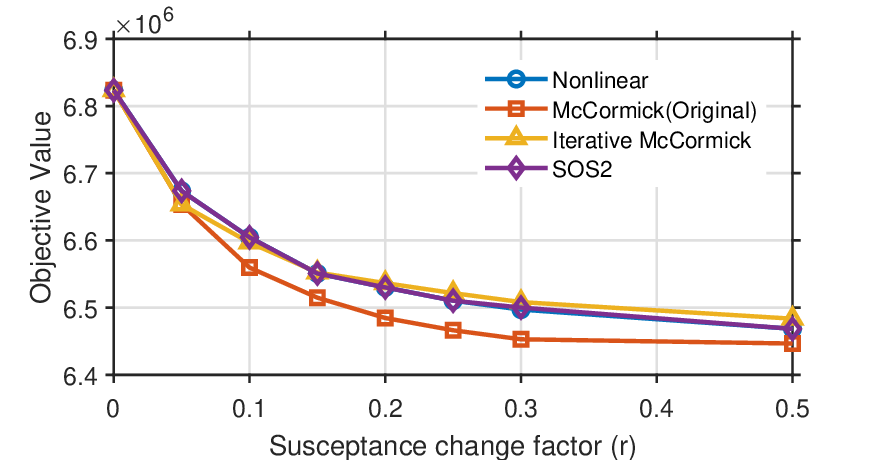}
    \vspace{-0.5em}
    \caption{Performance comparison of objective function costs under different methods for \texttt{Case300} using the nominal topology.}
    \label{fig:300_cost_Original}
    % \vspace{-1em}
\end{figure}
\begin{table}[h!]
% \vspace{-0.5em}
\centering
\caption{Error and computation time of relaxation/approx methods for different $r$ (Case300, Nominal Topology)}
 \vspace{-0.8em}
\label{table:300_original_accuracy}
\resizebox{\columnwidth}{!}{%
\begin{tabular}{c c cc cc cc}
\toprule
\multirow{2}{*}{$r$ (\%)} &
\multicolumn{1}{c}{Nonlinear} &
\multicolumn{2}{c}{McCormick (Original)} &
\multicolumn{2}{c}{Iterative McCormick} &
\multicolumn{2}{c}{SOS2} \\
% \cmidrule(lr){2-3} 
\cmidrule(lr){3-4} 
\cmidrule(lr){5-6} 
\cmidrule(lr){7-8}
& Time (s) 
 & Error (\%) & Time (s)
 & Error (\%) & Time (s)
 & Error (\%) & Time (s) \\
\midrule
0   & 1.18 & 0.000   & 1.05 & 0.000   & 1.06 & 0.000 & 1.80 \\
5   & 1.76 & 0.300  & 1.11 & 0.300  & {1.07} & {0.001} & 6.78 \\
10  & {1.91} & 0.681  & 1.10 & 0.117  & 2.13 & {0.000 }& 12.42 \\
15  & {1.99} & 0.550  & 1.11 & 0.021   & 3.19 & {0.000} & 8.40 \\
20 &  {2.07} & 0.697  & 1.10 & 0.099   & 4.21 & {0.000} & 31.03 \\
25  &  {2.16 }& 0.677  & 1.11 & 0.173   & 5.24 & {0.000} & 31.82 \\
30  &  {2.07} & 0.680  & 1.10 & 0.172   & 6.34 & {0.047} & 6.98 \\
50  &  {1.95} & 0.338  & 1.11 & 0.234   & 10.59 & {0.000 }& 9.53 \\
\bottomrule
\end{tabular}}
% \vspace{-1em}
\end{table}
\begin{figure}[!htbp]
    \vspace{-0.5em}
    \centering
    \includegraphics[width=0.88\linewidth]{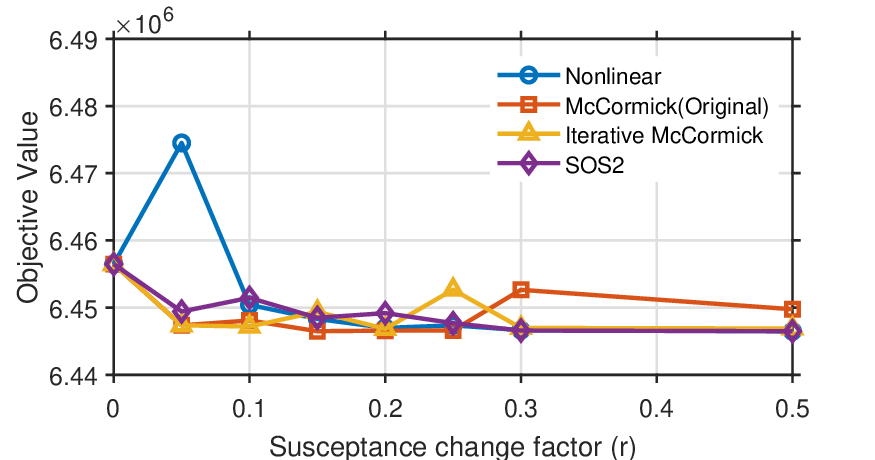}
    \vspace{-0.5em}
    \caption{Performance comparison of objective function costs under different methods for \texttt{Case300} using the optimal topology via NTO.}
    \label{fig:300_cost}
     \vspace{0.5em}
\end{figure}

% \begin{table*}[t]
% \centering
% \caption{Error and Computation Time of Relaxation Methods for Different $r$ (Case300, NTO)}
% \label{table:300_nto_accuracy}
% \resizebox{\textwidth}{!}{%
% \begin{tabular}{c ccc cc cc cc cc}
% \toprule
% \multirow{2}{*}{$r$ (\%)} &
% \multicolumn{3}{c}{Nonlinear} &
% \multicolumn{2}{c}{McCormick (Original)} &
% \multicolumn{2}{c}{Iterative McCormick} &
% \multicolumn{2}{c}{SOS2} \\
% \cmidrule(lr){2-4}
% \cmidrule(lr){5-6}
% \cmidrule(lr){7-8}
% \cmidrule(lr){9-10}
%  & Objective & Error (\%) & Time (s)
%  & Error (\%) & Time (s)
%  & Error (\%) & Time (s)
%  & Error (\%) & Time (s) \\
% \midrule
% 0  & 6{,}456{,}470.26\textsuperscript{$\dagger$} & 0.000 & 1205.93
%    & 0.000 & 1205.33
%    & 0.000 & 1205.69
%    & 0.000 & 1205.54 \\
% 10 & 6{,}450{,}406.29 & 0.000 & 920.84
%    & -0.036 & 41.42
%    & -0.050 & 128.78
%    & 0.016 & 47.61 \\
% 20 & 6{,}446{,}978.19 & 0.000 & 375.47
%    & -0.006 & 22.66
%    & -0.002 & 246.85
%    & 0.034 & 135.91 \\
% 30 & 6{,}446{,}644.38 & 0.000 & 342.35
%    & 0.093 & 20.10
%    & 0.005 & 365.00
%    & -0.001 & 75.62 \\
% 50 & 6{,}446{,}481.88 & 0.000 & 151.38
%    & 0.051 & 32.76
%    & 0.006 & 544.09
%    & -0.000 & 44.16 \\
% \bottomrule
% \end{tabular}
% }
% \\[1mm]
% \footnotesize \textsuperscript{$\dagger$}For $r=0\%$, all methods reached the time limit (1200 s); the reported objective values correspond to the best incumbent solutions at termination.
% \end{table*}
\begin{table}[h!]
\vspace{-1em}
\centering
\caption{Error and computation time of relaxation/approx methods for different $r$ (Case300, Optimal Topology using NTO)}
 \vspace{-0.8em}
\label{table:300_nto_small}
\resizebox{\columnwidth}{!}{%
\begin{tabular}{c c cc cc cc}
\toprule
\multirow{2}{*}{$r$ (\%)} &
\multicolumn{1}{c}{Nonlinear} &
\multicolumn{2}{c}{McCormick (Original)} &
\multicolumn{2}{c}{Iterative McCormick} &
\multicolumn{2}{c}{SOS2} \\
% \cmidrule(lr){2-3} 
\cmidrule(lr){3-4} 
\cmidrule(lr){5-6} 
\cmidrule(lr){7-8}
 & Time (s) 
 & Error (\%) & Time (s)
 & Error (\%) & Time (s)
 & Error (\%) & Time (s) \\
\midrule

0   & 1205.93\textsuperscript{$\dagger$}
   & 0.000 & 1205.33\textsuperscript{$\dagger$}
   & 0.000 & 1205.69\textsuperscript{$\dagger$}
   & 0.000 & 1205.54\textsuperscript{$\dagger$} \\
   
5  & 1206.95\textsuperscript{$\dagger$}
   & 0.419 & 81.97
   & 0.419 & 78.55
   & 0.388 & 179.80 \\ 
   
10 & 920.84
  & 0.036 & 41.42
  & 0.050 & 128.78
  & 0.016 &  47.61\\
15 & 564.82
   & 0.028 & 15.53
   & 0.017 & 185.17
   & 0.003 &  46.16\\
20 & 375.47
   & 0.006 & 22.66
   & 0.002 & 246.85
   & 0.034 & 135.91 \\
25 & 435.44
   & 0.012 & 22.39
   & 0.082 &312.15
   & 0.005 &  112.66\\

30  & 342.35
   & 0.093 & 20.10
   & 0.005 & 365.00
   & 0.001 & 75.62 \\

50  & 151.38
   & 0.051& 32.76
   & 0.006 & 544.09
   & 0.000 &  44.16\\
\bottomrule
\end{tabular}
}
\\
\footnotesize\textsuperscript{$\dagger$}Indicates instances that reached the 1200\,s time limit. At $r=0\%$, all methods reached the limit; at $r=5\%$, only the nonlinear formulation did.
% \vspace{-1em}
\end{table}
\begin{table*}[t]
\centering
\caption{Objective values, error (in parentheses, relative to the nonlinear objective value), \\ and computation time for larger testcases using different methods ($r = 10\%$).}
 \vspace{-0.8em}
\label{tab:obj_error_10}
\resizebox{\textwidth}{!}{%
\begin{tabular}{l
                cccc
                cccc}
\toprule
\multirow{2}{*}{IEEE Test bus system} &
\multicolumn{4}{c}{Objective Values (\$)} &
\multicolumn{4}{c}{Computation Time (s)} \\ 
\cmidrule(lr){2-5}
\cmidrule(lr){6-9}
& Nonlinear & McCormick (Original) & Iterative McCormick & SOS2
& Nonlinear & McCormick (Original) & Iterative McCormick & SOS2 \\
\midrule

\texttt{case300} &
6,604,576 & 6,559,597 (0.681\%) & 6,596,829 (0.117\%) & 6,604,576 (0.000\%) &
1.91 & 1.10 & 2.13 & 12.42 \\

\texttt{case588\_sdet} &
1,787,169 & 1,786,860 (0.017\%) & 1,787,144 (0.001\%) & 1,787,413 (0.014\%) &
59.46 & 2.29 & 3.79 & 5.14 \\

\texttt{case1354pegase} &
25,628,319 & 25,581,370 (0.183\%)  & 25,620,519 (0.03\%) & 25,635,474 (0.028\%) &
3473.10 & 5.07 & 10.22 & 24.51 \\

\texttt{case1888rte} &
19,392,895 & 18,545,441 (4.370\%)  & 18,701,887 (3.563\%) & 18,729,993 (3.42\%\textsuperscript{$\dagger$}) &
27850.38 & 7.85 & 16.08 & 14400 \\

\bottomrule
\end{tabular}}
\footnotesize\textsuperscript{$\dagger$}Indicates solution up to 0.26\% gap interrupted due to time limit: 4 hours.
\vspace{-2em}
\end{table*}

The optimized objectives using the nonlinear, McCormick, iterative-McCormick, and SOS2 are shown in Fig.~\ref{fig:300_cost_Original}. From the plot, we can make two observations. First, the cost goes down with an increase in $r$ i.e., the flexibility in changing the line susceptance thanks to VIDs. Second, the iterative McCormick and SOS2 costs are quite close to the nonlinear case, whereas the original McCormick deviates from the nonlinear cost as $r$ increases. This happens because an increase in $r$ leads to a larger McCormick relaxation and therefore it is more inexact. 

We also quantify the difference between the costs obtained by relaxed and approximated methods with respect to the nonlinear method in Table~\ref{table:300_original_accuracy}. It also shows the computation time for each method. It shows that the iterative McCormick method improves upon the error, but it takes more time due to the iterations. For this case, SOS2 performs the best both in terms of time and accuracy.

The results with optimized topology, i.e., solving the NTO problem with VIDs, are shown in Fig~\ref{fig:300_cost}. Here, we see that the cost does not change much with an increase in $r$, this is because NTO already reduced the cost by a large margin, and there is not much improvement that can be achieved using VIDs for this testcase.
% on reducing the cost function value is minor as the $r$ value increases compared to when NTO is not applied. 
In some cases, the value of the cost function increases as the $r$ increases. This is because the Gurobi solver's Gap was set to 0.1\%, leading to errors. The error is especially large when r is 5\% between the Nonlinear model and the other models. This is caused by the solver not converging to a solution within the time limit (1200 seconds). The error and computation time are presented in Table~\ref{table:300_nto_small}. We again observed that SOS2 is the best-performing approximation.

\subsubsection{Larger testcases}
To evaluate whether the proposed scheme is scalable on larger networks, we simulate it for \texttt{\detokenize{Case588_sdet}}, \texttt{Case1354pegase}, and \texttt{Case1888rte}.
For brevity, we present the result for $r = 10\%$ considering the nominal topology. The results are presented in Table~\ref{tab:obj_error_10} comparing the cost, computation time, and accuracy. We observe that the proposed iterative McCormick achieves the best performance in accuracy and computation time for all the larger systems.

\section{Conclusions}
This work considered the problem of optimizing the operation of distributed variable impedance devices with network topology optimization in power transmission networks. Due to the varying nature of the line reactances, this problem contained bilinear constraints. The work proposed using McCormick envelopes to relax the bilinear constraints, which results in a large error compared to the objective obtained by the nonlinear model. Therefore, we proposed an iterative approach to apply McCormick envelopes in small steps. The proposed method was compared against the nonlinear method, the original McCormick relaxation, and the SOS2 approximation. The comparison was performed with respect to the error against the nonlinear method and the computation time. We observe that the proposed method performs the best on large systems such as \texttt{\detokenize{Case588_sdet}}, \texttt{Case1354pegase}, and \texttt{Case1888rte}, whereas for the smaller networks, e.g., \texttt{case300}, SOS2 performs the best.
Future work will focus on using the proposed scheme for optimal planning of variable impedance devices in the transmission system.
\label{sec:conclusion}

% \vspace{-0.5em}
\bibliographystyle{IEEEtran}
\bibliography{bibliography.bib}
\end{document}